\documentclass{article}

\usepackage[utf8]{inputenc}
\usepackage{amsmath}
\usepackage{amssymb}
\usepackage{theorem}

\newtheorem{Def}{Definition}
\newtheorem{Theo}{Theorem}
\newtheorem{Prop}[Theo]{Proposition}

\newtheorem{Cor}{Corollary}[Theo]

\newenvironment{Proof}[1][Proof]{\paragraph{{#1}}}%
                {{\hfill\(\Box\)\\}}
                {{\hfill\(\Box\)\\}}
        {\paragraph{{#1}}\begin{list}{}{}\item
        }{\end{list}}

\newcommand{\cand}{\text{ and }}

\newcommand{\coll}[1]{\ensuremath{\left\{ {#1}\right\} }}

\newcommand{\paren}[1]{\ensuremath{\left( {#1} \right)}}

\newcommand{\set}[2]{\ensuremath{\left\{\left.#1\,\,\vphantom{#2}\right|\,#2\right\}}}
\newcommand{\fall}[1]{{\forall\,{#1},\ }}

\newcommand{\mc}[1]{{\mathcal{#1}}}

\newcommand{\mb}[1]{{\bf #1}}

\title{A priori Knowledge and \\ the Kochen-Specker Theorem}
\author{Olivier Brunet\footnote{olibrunet@free.fr}}

\begin{document}

\maketitle

\begin{abstract}
We introduce and formalize a notion of “a priori knowledge” about a quantum system, and show some properties about this form of knowledge. Finally, we show that the Kochen-Specker theorem follows directly from this study.
\end{abstract}

\section{Introduction}

In this article, we introduce a form of “a priori knowledge” about a quantum system, formalized by what we call Sasaki filters of an orthomodular lattice. We then show that in the context of a Hilbert space of dimension at least 3, the Sasaki filters of the associated Hilbert lattice cannot contain more than one atom. This property is shown to imply the Kochen-Specker theorem \cite{KochenSpecker67} and also provides arguments against some weaker forms of value-definiteness.

In the next section, we introduce our notion of “a priori knowledge” and exhibit the properties we want this notion to convey. In the next section, we provide a mathematical formalization of this notion, and we introduce some notations and basic properties. Then, in section \ref{Sec:Theorem}, we focus on the Hilbert lattice of a Hilbert space of dimensation at least 3, we present a theorem about Sasaki filters in this context, and show how it implies the Kochen-Specker theorem.

\section{A priori Knowledge about Quantum Systems}

Let us consider a quantum system $\mc S$, and suppose that a measurement $m$ is performed on it. The result of this measurement can be represented by an eigenspace $E$ of the hermitian operator corresponding to $m$.

Now, if moreover we consider that after $m$ has been performed, the state of $\mc S$ has not changed, then the following statement has to be considered as valid:

\smallskip

\centerline{“if we perform measurement $m$ on $\mc S$, the outcome will be $E$”}

\smallskip

Such a statement corresponds to what we call {\em a priori knowledge} about the system: it provides informations about the state of the system, and is {\em a priori} in the sense that it tells what the result of a not-yet-performed measurement would be. Finally, the term {\em knowledge} refers to the fact that it is a non-probabilistic statement but rather a testable logical assertion about the system.

\medskip

Let us get back to the previous situation, where one has some priori knowledge about $\mc S$ provided by the previous statement: “if we perform measurement $m$ on $\mc S$, the outcome will be $E$”. Suppose now that one can also perform another measurement $m'$ of $\mc S$ which corresponding hermitian operator also has $E$ as an eigenspace. If $m'$ is performed on $S$, then quantum mechanics tells us that the outcome would also be $E$.

This means that, more generally, a statement like “if we perform measurement $m$ on $\mc S$, the outcome will be $E$” implies that “performing on $\mc S$ any measurement with $E$ as a possible outcome (i.e. such that $E$ is an eigenspace of the associated hermitian operator) would also yield $E$ as the outcome”. As a consequence, no reference to a particular measurement has to be made as in the former statement and in the following, we will only use statements of the latter form, expressed in a more compact way as:

\smallskip

\centerline{“$\mc S$ verifies $E$”}

\smallskip

This form of statement emphasizes the fact that the kind of a priori knowledge we deal with only concerns eigenspaces of hermitian operators, that is more generally closed subspaces of the Hilbert space $\mc H$ used to model our system.

\medskip

From the previous discussion, it is clear that the statement “$\mc S$ verifies $E$” means that the state of $\mc S$ actually lies in $E$. This remarks leads to two properties concerning the subspaces verified by a system:
\begin{enumerate}
\item Let $E$ and $F$ be two subspaces such that $E \subseteq F$. It is clear then that if “$\mc S$ verifies $E$”, then “$\mc S$ verifies $F$”.
\item Let $E$ and $F$ be two compatible subspaces such that “$\mc S$ verifies $E$” and ”$\mc  S$ verifies $F$”. Then, as a consequence, “$\mc S$ verifies $E \cap F$”.
\end{enumerate}
This suggests that in order to study the notion of a priori knowledge that we have just defined, we should consider collections of subspaces that verify these two properties. It is precisely this type of collection which we will formalize in the next section.

\section{Sasaki Filters}

In order to formalize what we call a {\em Sasaki filter}, we first introduce a few notions and definitions.

\begin{Def}[Upward-closed subset]
Given a poset $\mc P$, a upward-closed subset $S$ of $\mc P$ is an element of $\mc P$ such that~:
$$ \fall {x \in S} \fall {y \in \mc P} (x \leq y \Rightarrow y \in S) $$
Let $\wp^\uparrow(\mc P)$ denote the set of upward-closed subsets of $\mc P$.
\end{Def}

Given an element $x \in \mc P$, $x^\uparrow$ will denote the set $\set{y \in \mc P}{x \leq y}$. Obviously, $x^\uparrow \in \wp^\uparrow(\mc P)$. Also, it is worth noting that $\wp^\uparrow(\mc P)$ is stable by arbitrary unions and intersections and as the structure of a complete lattice.

\medskip

We now focus on the use of orthomodular lattices, where one can define a binary operation, the {\em Sasaki projection} in the following way~: $x \,\&\, y = y \wedge (x \vee y^\bot)$. This operation is important in the field of quantum logic, since while orthomodular lattices are an convenient generalization of Hilbert lattices, the Sasaki projection is the corresponding generalization of orthogonal projection.

In the following, we will show that the Sasaki projection is also particularly important in relation to a priori knowledge.

\begin{Def}[Sasaki filter]
Given an orthomodular lattice $\mc L$, a Sasaki filter $F$ of $\mc L$ is an upward-closed element of $\mc L$ such that~:
$$ \fall {x,y \in F} x \, \& \, y \in F $$
A Sasaki filter $F$ is said to be {\em proper} if it is non-empty and different from $\mc L$. Finally, let $\wp^\uparrow_\&(\mc L)$ denote the set of Sasaki filters of $\mc L$.
\end{Def}

Given an element $x$ of $\mc L$, $x^\uparrow$ is in $\wp^\uparrow_\&(\mc L)$. Moreover, even though $\wp^\uparrow_\&$ is not stable by arbitrary union, it is stable by arbitrary intersection, and can also be given the structure of a complete lattice.

\smallskip

The next proposition shows that Sasaki filters constitute a formalization of the notion of a priori knowledge which we introduced in the previous section. 

\begin{Prop} \label{Prop:StableMeet}
An upward-closed subset $S$ of an orthomodular lattice $\mc L$ is a Sasaki filter if and only if it is stable by compatible meet.
\end{Prop}
\begin{Proof}
First, suppose that $S$ is a Sasaki filter and let $x$ and $y$ be two compatible elements of $S$. Then one has $x \wedge y = x \,\&\, y$ so that $x \wedge y$ is in $S$.

Conversely, let $x$ and $y$ be two elements of $S$. One has $x \,\&\,y = y \wedge (x \vee y^\bot)$. Since $x \leq x \vee y^\bot$, it follows that $x \vee y^\bot$ is in $S$. But as $y$ and $x \vee y^\bot$ are compatible, their meet $x\,\&\, y$ is also in $S$.
\end{Proof}

Finally, we introduce some notations concerning the definition of a Sasaki projection-like operation for upward-closed subsets.

\begin{Def}
Given an orthomodular lattice $\mc L$, we define $\overline\& : \wp^\uparrow(\mc L) \rightarrow \wp^\uparrow(\mc L)$ as:
$$ \overline\&(S)=\bigcup \set{(x\,\&\,y)^\uparrow}{x,y \in S} $$
Moreover, we introduce the following notations:
$$ \overline\&^n(S)=\underbrace{\overline\& \circ \cdots \circ \overline\&}_{\hbox{$n$ times}}(S) \qquad \overline\&^\infty(S)=\bigcup_{n \in \mb N} \overline\&^n(S) $$
\end{Def}

Here are a few easy facts about $\overline\&$:
\begin{Prop}
For all $S \in \wp^\uparrow(\mc L)$, one has:
\begin{gather*}
S \subseteq \overline\&(S) \subseteq \overline\&^2(S) \subseteq \cdots \subseteq \overline\&^n(S) \subseteq \cdots \subseteq \overline\&^\infty(S) \\
\overline\&^\infty(S) \in \wp^\uparrow_\&(\mc L) \qquad \overline\&^\infty(S) = \bigcap \set {S' \in \wp^\uparrow_\&(\mc L)}{S \subseteq S'}
\end{gather*}
\end{Prop}

\section{A Theorem about Sasaki Filters} \label{Sec:Theorem}

\subsection{A Geometric Lemma}

Let us consider the  Hilbert space $\mb R^3$, and let $u$ and $v$ denote two non-nul vectors such that $u\cdot v >0$. In an appropriate orthonormal basis $\mb e=\coll{e_1,e_2,e_3}$, one can write~:
$$ u = \left( \begin{array}{c} 1\\0\\0 \end{array} \right) \qquad v = \left( \begin{array}{c}
\cos \theta \\ \sin \theta \\ 0 \end{array} \right) $$
with $\theta \in ]0,\frac\pi2[$.

In the same basis, given a real $\varphi$, let us introduce $w_\varphi = \left(\begin{array}{c}
0 \\ \cos \varphi \\ \sin \varphi \end{array} \right) $.

Let $E_\varphi$ be the plan spanned by $u$ and $w_\varphi$~: $E_\varphi=\hbox{span}\coll{u,w_\varphi}$ and let $\pi_\varphi(v)$ denote the orthogonal projection of $v$ on $E_\varphi$. Finally, let $v_\varphi$ denote this projection after normalization~:

$$ v_\varphi = \frac{\pi_\varphi(v)}{||\pi_\varphi(v)||}$$

Simple calculations show that~:

$$ v_\varphi=\frac1{\sqrt{\cos^2\theta+\sin^2\theta\cos^2\varphi}} \left( \begin{array}{c} \cos \theta \\ \sin \theta \cos^2 \varphi \\ \sin \theta \cos \varphi \sin \varphi \end{array} \right) $$

As a consequence, one has~:
$$ v_\varphi \cdot v_\psi = \frac {\cos^2 \theta+\sin^2 \theta \paren {\cos^2 \varphi \cos^2\psi + \cos \varphi \cos \psi \sin \varphi \sin \psi } } {\sqrt{\paren{\cos^2\theta+\sin^2\theta\cos^2\varphi}\paren{\cos^2\theta+\sin^2\theta\cos^2\psi}}}$$

\begin{Prop}[Geometric Lemma]
With the previous notations, if $0<\theta<\frac\pi2$, one has: 
$$ \set{v_\varphi \cdot v_\psi}{\varphi, \psi \in [0;2\pi]}=\left[\frac{3\cos\theta-1}{\cos\theta+1};1\right]$$
\end{Prop}
\begin{Proof}
First, it is clear that the function $(\varphi, \psi) \mapsto v_\varphi\cdot v_\psi$ is continuous, so that the set on the left-hand side of the equality has to be an interval.
If $\varphi=\psi$, then obviously, $v_\varphi \cdot v_\psi = 1$. Now, if~:
$$\varphi = \arccos \sqrt{\frac{\cos \theta}{1+\cos \theta}} \quad \hbox{and} \quad \psi = - \arccos \sqrt{\frac{\cos \theta}{1+\cos \theta}}\hbox{,}$$
one obtains $v_\varphi \cdot v_\psi = \cfrac{3\cos\theta-1}{\cos\theta+1}$. Thus, we have shown that~:
$$ \left[\frac{3\cos\theta-1}{\cos\theta+1};1\right] \subseteq \set{v_\varphi \cdot v_\psi}{\varphi, \psi \in [0;2\pi]}$$
The equality is obtained by studying the extremas of $(\varphi, \psi) \mapsto v_\varphi\cdot v_\psi$, which are attained either for $\varphi \equiv \psi\,[2\pi]$ or for $\varphi \equiv -\psi\,[\pi]$ and~:
$$ \sin \varphi=0 \quad \hbox{or}\quad \cos\varphi=0 \quad \hbox{or}\quad \sin^2\theta \cos^4\varphi+2\cos^2\theta\cos^2\varphi-\cos^2\theta=0\hbox{.} $$
\end{Proof}

Let us define $f$ on $[0,1]$ by $f(x)=\frac{3x-1}{x+1}$. It is easy to verify that $f(0)=-1$, $f(\frac13)=0$, $f(1)=1$, $f$ is strictly increasing and that $f(x)<x$. Moreover, let us define a sequence $(c_n)$ by~:
$$ \left\{\begin{array}{l} c_0 = 0 \\ c_{n+1} = f^{-1}(c_n) = \cfrac{1+c_n}{3-c_n} \end{array} \right. $$
From its definition, it appears that $(c_n)$ is an homographic sequence and one can express $c_n$ as a function of $n$~:
$$\fall {n \in \mb N} c_n = \cfrac n {n+2}\hbox{.}$$
Finally, we define $(\theta_n)$ by~:
$$\fall {n \in \mb N} \theta_n = \arccos\paren{c_n} = \arccos\Bigl(\cfrac n {n+2}\,\Bigr)\hbox{.}$$
Clearly, $\theta_0=\frac\pi2$ and $\lim\limits_{n\rightarrow\infty} \theta_n=0$.

\subsection{The Main Result}

In the following, we will consider Hilbert spaces over $\mb R$, $\mb C$ or $\mb H$. It is well known that the Hilbert lattice (i.e. the set of closed subspaces) of a Hilbert space $\mc H$, denoted $\mc L_{\mc H}$, is an orthomodular lattice, as the study of these structures have motivated the development of the field of quantum logic.

\ 

When considering Hilbert spaces and their related Hilbert lattices, the relationship between orthogonal projection and Sasaki projection is illustrated by this proposition:

\begin{Prop}
Given a Hilbert space $\mc H$, one has:
$$ \fall{A,B \in \mc L_{\mc H}} A \,\&\, B = \set{\Pi_B(u)}{u \in A} $$
\end{Prop}

Let us now focus on the Sasaki filters of $\mc L_{\mc H}$ where $\mc H$ is a Hilbert space of dimension at least $3$. Our goal is to show that these Sasaki filters cannot contain two or more atoms of $\mc L_{\mc H}$. The next proposition provides the induction tool to prove this result, where we use the following notation~: given two atom $A=\hbox{span}(u)$ and $B=\hbox{span}(v)$, we define $d(A,B)$ as~:
$$ d(A,B) = \frac{|u \cdot v|}{||u||\,||v||}$$
This notation makes sense, as the value of $d(A,B)$ does not depend on the choice of $u$ and $v$.

\begin{Prop}\label{Prop:Induction}
Let $\mc H$ be a Hilbert space of dimension at least $3$. Let $A$ and $B$ be two atoms of $\mc L_{\mc H}$ and $n$ an integer such that $d(A,B) \geq \theta_n$. Then there exists two atoms $A'$ and $B'$ in $\overline\&(A^\uparrow \cup B^\uparrow)$ such that $d(A',B')=\theta_{n-1}$.
\end{Prop}
\begin{Proof}
Let $u$ and $u$ be two normailzed vectors of $\mc H$ such that~:
$$ A = \hbox{span}\{u\} \qquad B = \hbox{span}\{v\} \qquad u\cdot v = \cos d(A,B) $$
Using previous notations, it is moreover possible to define an orthonormal basis $\mb e$ of $\mc H$ such that~:
$$ u = e_1 \qquad v = \cos d(A,B) \, e_1 + \sin d(A,B) \, e_2 $$
Now, for every $\varphi \in \mb R$, let us define as previously~: 
$$w_\varphi = \cos \varphi \, e_2 + \sin \varphi \, e_3 \qquad E_\varphi = \hbox{span}\{u,w_\varphi\} \qquad v_\varphi = \cfrac{\Pi_{E_\varphi}(v)}{||\Pi_{E_\varphi}(v)||} $$
As mentionned before, $\hbox{span}\{v_\varphi\} = \hbox{span}\{\Pi_{E_\varphi}(v)\} = \hbox{span}\{v\}\,\&\,E_\varphi = B \,\&\, E_\varphi$. Now, since $A \subseteq E_\varphi$, one has $E_\varphi \in A^\uparrow \cup B^\uparrow$ so that~:
$$\hbox{span}\{v_\varphi\} \in \overline\&(A^\uparrow \cup B^\uparrow)$$
From the hypothesis that $\cos d(A,B) \leq \cos(\theta_n)$ and the monotony of $f$, one~has:
$$ \cfrac{3\cos d(A,B)-1}{\cos d(A,B)+1} = f\bigl(\cos d(A,B)\bigr) \leq f\bigl(\cos \theta_n) = \cos \theta_{n-1} $$
Using our Geometric Lemma, this implies that there exists two real numbers $\alpha$ and $\beta$ such that $v_\alpha \cdot v_\beta = \cos \theta_{n-1}$.

With $A'=\hbox{span}\{v_\alpha\} = B \,\&\,E_\alpha$ and $B'=B \,\&\,E_\beta$, one finally has~:
$$\{A',B'\} \subseteq \overline\&(A^\uparrow \cup B^\uparrow) \qquad \hbox{and} \qquad d(A',B')=\theta_{n-1}$$
\end{Proof}

We now turn to our main result:

\begin{Theo} \label{Theo:Main}
Given a Hilbert space $\mc H$ of dimension at least $3$, every proper Sasaki-filter $F$ of $\mc L_{\mc H}$ contains at most one atom.
\end{Theo}
\begin{Proof}
Suppose that $F$ contains two distinct atoms $A$ and $B$. One has $d(A,B) > 0$ so that there exists an integer $n$ such that $\theta_n \leq d(A,B)$.

Using proposition \ref{Prop:Induction}, there exists two atoms $A_1$ and $B_1$ in $\overline\&(A^\uparrow \cup B^\uparrow)$ such that $d(A_1,B_1)=\theta_{n-1}$ and by induction, there exists two atoms $A_n$ and $B_n$ in $\overline\&^n(a^\uparrow \cup b^\uparrow)$ such that $d(A_n,B_n)=\theta_{0}=0$.

This implies that $A_n \leq B_n^\bot$, so that~:
$$ \bot = A_n \,\&\, B_n \in \overline\&^{n+1}(A^\uparrow \cup B^\uparrow) \subseteq F $$
This is not possible since $F$ is a proper Sasaki ideal of $\mc L_{\mc H}$. As a consequence, $F$ contains at most one atom.
\end{Proof}

\begin{Cor}
Given a Hilbert space $\mc H$ of dimension at least $3$, every proper Sasaki-filter $F$ of $\mc L_{\mc H}$ is such that if $F$ contains an atom $A$ of $\mc L_{\mc H}$, then $F=A^\uparrow$.
\end{Cor}
\begin{Proof}
Since $F$ contains an atom $A$, one has $A^\uparrow \subseteq F$. Conversely, for every $E \in F$, $A \,\&\, E$ is an atom and belongs to $F$. This implies using the previous theorem that $ A\,\&\,E = A$ which is equivalent to $A \leq E$. Thus, we have shown that $F \subseteq A^\uparrow$.
\end{Proof}

\subsection{A Proof of the Kochen-Specker Theorem}

The Kochen-Specker theorem \cite{KochenSpecker67} is a very important result about the possibility of hidden variables theories of quantum mechanics. It asserts that there exists sets of observables which cannot be assigned values simultaneously in a consistent way. One of the simplest ways to state this theorem is to assert that there is no two-valued measures (or valuations) on the Hilbert lattice of a Hilbert space of dimension at least $3$. This theorem is also a central result in showing that the underlying logic on quantum mechanics cannot be understood in terms of classical logic, except at the cost of most logical operations \cite{Svozil98Book,Calude99Embedding}.

In order to precisely show how the Kochen-Specker follows from theorem \ref{Theo:Main}, we introduce the notion of a pre-valuation which is a generalization of valuations. We then show that the latter theorem is equivalent to stating that in dimension at least 3, pre-valuations can contain at most one atom, so that there are not valuations.

\begin{Def}[Pre-valuation]
A {\em pre-valuation} on an orthomodular lattice $\mc L$ is a function $\nu : \mc L \rightarrow \coll{0,1}$ which~verifies:
\begin{gather*}
\nu(\top)=1 \\
\fall {x,y \in \mc L} \bigl( x \,\bot\, y \Rightarrow \nu(x \vee y) \geq \nu(x)+\nu(y) \bigr) \\
\fall{x,y \in \mc L} \bigl( x \hbox{\ and\ } y \hbox{\ compatible} \Rightarrow \nu(x \wedge y) = \nu(x) \times \nu(y) \bigr)
\end{gather*}
\end{Def}

From this definition, it is clear that any valuation (i.e. two-valued measure) on $\mc L$ is a pre-valuation on $\mc L$.

\begin{Prop}\label{Prop:KSLemma}
Given an orthomodular lattice $\mc L$, a function $\nu: \mc L \rightarrow \coll{0,1}$ is a pre-valuation if and only if the set $ \set{x \in \mc L}{\nu(x)=1}$ is a proper Sasaki filter of $\mc L$.
\end{Prop}
\begin{Proof} This is a direct consequence of \ref{Prop:StableMeet} since the definition of a pre-valuation can be restated as:
\begin{gather*}
\nu(\top)=1 \\
\fall {x,y \in \mc L} \paren{x \leq y \cand \nu(x)=1} \Rightarrow \nu(y)=1 \\
\fall {x,y \in \mc L} x \hbox{ and } y \hbox{ compatible} \Rightarrow \nu(x \wedge y) = \nu(x) \times \nu(y) \\
\end{gather*}
\end{Proof}

This proposition shows that prevaluations are actually another way to represent Sasaki filters. Thus, theorem \ref{Theo:Main} can be equivalently rephrased as: “on a Hilbert space of dimension at least $3$, for every pre-valuation, at most one atom evaluates to $1$”.

And this statement directly implies the Kochen-Specker theorem:

\begin{Theo}[Kochen-Specker]
Given a Hilbert space $\mc H$ of dimension at least $3$, there is no valuation on $\mc L_{\mc H}$.
\end{Theo}
\begin{Proof}
This is a direct consequence of theorem \ref{Theo:Main}.
\end{Proof}

\section{Conclusion and perspective}

We have introduced a notion of “a priori knowledge” about a quantum system, formalized by Sasaki filters, i.e. collections of properties (represented by closed subspaces of a Hilbert space, and more generally by elements of an orthomodular lattice) which is stable by logical implication and compatible conjunction. 

In the case of the Hilbert lattice corresponding to a Hilbert space of dimension at least 3, we have shown that Sasaki filters can not contain more than one atom, which implies the Kochen-Specker theorem.

However, this result also appears to be an strong argument against some forms of partial value-definiteness, since in terms of sphere coloring, this result shows that at most a single point of the sphere can be attributed a color that is different from others. It would also be interesting to consider the main result of this article with regards to the impossibility of performing infinite precision measurements. 

To that respect, it seems important to pursue the study of Sasaki filters in at least two directions: first, to explore the structure of $\mc P^\uparrow_\&(\mc L)$ more in details, and second, to explicit the role that Sasaki filters could play as a description of the state of a quantum state.

Finally, it should be noticed that this result concerning the Kochen-Specker theorem can be closely related to a similar result by J.~D.~Malley for the Bell-Kochen-Specker theorem \cite{Malley06Bell}.


\begin{thebibliography}{}

\bibitem[Calude et~al., 1999]{Calude99Embedding}
Calude, C.~S., Hertling, P.~H., and Svozil, K. (1999).
\newblock Embedding quantum universes in classical ones.
\newblock {\em Foundations of Physics}, 29(3):349--390.

\bibitem[Kochen and Specker, 1967]{KochenSpecker67}
Kochen, S. and Specker, E.~P. (1967).
\newblock The problem of hidden variables in quantum mechanics.
\newblock {\em Journal of Mathematics and Mechanics}, 17(1):59--87.

\bibitem[Malley, 2006]{Malley06Bell}
Malley, J.~D. (2006).
\newblock The collapse of bell determinism.
\newblock {\em Physical Letters A}, 359(2):122--125.

\bibitem[Svozil, 1998]{Svozil98Book}
Svozil, K. (1998).
\newblock {\em Quantum Logic}.
\newblock Springer.

\end{thebibliography}

\end{document}